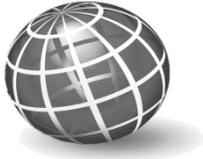

# The Journal of **Macro**Trends in Technology and Innovation

# Automatic Knot Adjustment Using Dolphin Echolocation Algorithm for B-Spline Curve Approximation

**Hasan Ali AKYÜREK\*, Erkan ÜLKER\*\*, Barış KOÇER\*\***
*\*Necmettin Erbakan University, School of Applied Sciences, Department of Management Information Sciences*
*\*\*Selcuk University, Faculty of Engineering, Department of Computer Engineering*

## Abstract

*In this paper, a new approach to solve the cubic B-spline curve fitting problem is presented based on a meta-heuristic algorithm called "dolphin echolocation". The method minimizes the proximity error value of the selected nodes that measured using the least squares method and the Euclidean distance method of the new curve generated by the reverse engineering. The results of the proposed method are compared with the genetic algorithm. As a result, this new method seems to be successful.*

Keywords: *B-spline Curve Approximation, Cubic B-spline, Data Parameterization on B-Spline, Dolphin Echolocation Algorithm, Knot adjustment*

1. INTRODUCTION

B-Spline curve fitting is a classical problem for computer aided geometric design [1]. For example, de facto for the CAD / CAM and related graphic design industries, and in most geometric modeling areas is that non-polynomial parametric curves should be transformed into non-uniform rational B-splines. Similarly, in vector font modeling problems, fonts are often fitted with a soft-pass B-Spline. In practical applications, the distance between the target curve and the fitted B-Spline curve must be less than a predetermined tolerance, and the resulting curve is called an error-bound approach. Euclidean distance method is used to measure the value corresponding to the distance between two curves.

## 2. B-Spline Curve Fitting Problem

The problem of B-Spline curve fitting is expressing the target curve with minimum tolerance through b-spline curves. The target curve can be two or three dimensional [2]. The scope of this paper

- The parameterization of the target data points
- The convergence of the minimum error tolerance with the B-Spline curves using the automatically placed minimum control point constitute.

A B-spline curve is expressed as equation (1).

$$P(u) = \sum_{i=0}^{n} p_i N_{i,p}(u) \ \#(1)$$

where $p_i$ is the $i$th control point and $N_{i,p}$ is the main function of B-Spline curves.

The main function $N_{i,p}$ of B-Spline curve for given knot vector $t$ with degree $p$ is expressed as equation (2).

$$N_{i,0}(u) = \begin{cases} 1, & t_i \leq u \leq t_{i+1} \\ 0, & otherwise \end{cases}$$

$$N_{i,p}(u) = \frac{u - t_i}{t_{i+p} - t_i} N_{i,p-1}(u) + \frac{t_{i+p+1} - u}{t_{i+p+1} - t_{i+1}} N_{i+1,p-1}(u) \ \#(2)$$

Further information on the B-Spline curves can be found [3].

### A. Methods on Data Parameterization

Because of B-Spline curves are parametric curves, the target data points need to be parameterized in the B-Spline curve fitting. However calculation of optimum data parameterization is theoretically quite difficult, different ways of data parameterization are used in applications. Three methods of uniform parameterization, chord-length parameterization, and centripetal parameterization are emerging in researches based on previous studies [4,5,6]. In this study centripetal parameterization method is used.

### B. Euclidean Distance Minimization

The Euclidean distance is used to calculate the error between the target curve and the B-Spline fitted curve. The Euclidean distance is calculated by an equation (3)

$$D = \sqrt{\sum_{i=1}^{L} \big(C(i) - B(i)\big)^2} \ \#(3)$$

where $C(i)$ is the $i$th data in original dataset, $B(i)$ is the $i$th data in the fitted curve. The general approach of this paper is to minimize this distance and express the B-Spline curve with minimum control point at the same time. Thus the Euclidean distance and the number of control points is treated together in the fitness function.

### C. Dolphin Echolocation Algorithm

The dolphin echolocation algorithm presented by Kaveh and Ferhoudi is an optimization algorithm that is inspired by the hunting principles of bottlenose dolphins through sonar waves [7]. The dolphins explore the entire search area for a specific effect to hunt. As they approach their prey, they try to focus on the target by limiting the number of waves they send by limiting their search. This algorithm implements search by reducing the distance to the target. The search space must be sorted before beginning to search. The alternatives of each variable to be optimized must be sorted in ascending or descending order. If these alternatives have more than one characteristic, they should be sorted according to the most important one. In the use of this technique, for example, for the variable $j$, the vector A$j$ in length LA$j$ forms the columns of the Alternatives Matrix. In addition, a convergence curve is used to change the convergence factor during the optimization process. The variation of this trend throughout the iterations is calculated by equation (4).

$$PP(Loop_i) = PP_1 + (1 - PP_i) * \frac{Loop_i^{Power} - 1}{(LoopsNumber)^{Power} - 1} \#(4)$$

where $PP$ is the probability of being pre-defined, $PP_1$ is randomly selected probability for the first iteration, $Loop_i$ is the number of the current iteration, $Power$ is the rank of the curve, and $LoopsNumber$ is the total number of iterations. Algorithm requires a location matrix $L_{NL*NV}$ in the variable number $NV$ at the location count $NL$.

The main steps of Dolphin Echolocation (DE) for discrete optimization are as follows:
1. Create $NL$ locations randomly for dolphin.
2. $PP$ of current iteration is calculated using the equation (4).
3. Fitness is calculated for every location.
4. Calculate cumulative fitness according to the following dolphin rules

(a)

for i = 1 to the number of locations

for j = 1 to the number of variables

find the position of L(i,j) in jth column of the Alternatives matrix and name it as A.

for k = -Re to Re

where $AF(A+k)_j$ is the cumulative fitness of the selected alternative $(A+k)$ for the $j$th variable (the numbering of the alternatives is the same as the ordering of the alternative matrix); $Re$ is the diameter of the influence of the neighbor affected by the cumulative fitness of alternative *A*. It is recommended that this diameter should not be more than 1/4 of the search space; *Fitness(i)* is the fitness of the *i*th location. The fitness should be defined as the best answers will get higher value. In other words, the goal of optimization should be to maximize the fitness.

AF must be calculated using a reflective property by adding alternatives near the edges (if *A + k* is not valid, i.e. *A + k <0* or *A + k> LAj*).

In this case, if the distance of the alternative to the edge is small, the same alternatives appear in the mirror as if a mirror were placed on the edge.

(b) A small *ɛ́* value is added to *AF* sequences *AF=AF+ɛ́* in order to distribute probabilities uniformly in search space. Here *ɛ́* should be chosen according to the way of describing the fitness. The best choice is lower than the lowest fitness value achieved.

(c) Find the best location for this loop and call it as *"Best location"*. Find the alternatives assigned to the best location variables and set their *AF* to zero.

In another saying:

for j = 1: Number of variables

    for i = 1: Number of alternatives

      if i = The best location(j)

        $AF_{ij} = 0$

5. For the variable *j (j = 1 to NV)*, calculate the probability by choosing alternative *i (i = 1 to ALj)* according to the equation (6).

$$P_{ij} = \frac{AF_{ij}}{\sum_{i=1}^{LA_j} AF_{ij}} \#(6)$$

6. Assign *PP* probability to all alternatives of all selected variables for the best location and distribute the remaining probability to other alternatives according to the form below:

for j = 1 to Number of variables

    for i = 1 to Number of alternatives

      if i = The best location(j)

        $P_{ij} = PP$

7. Calculate the next step locations according to the assigned probability to each alternate. Repeat steps 2-6 maximum iteration number times.

3. AUTOMATIC KNOT ADJUSTMENT BY DOLPHIN ECHOLOCATION ALGORITHM

In the problem of B-Spline curve fitting, the fitted curve is tried to converge to the target curve with minimum tolerance and with minimum control point. In that case, such nodes must be selected for the given *N* points so that the error tolerance and the number of control points of the nearest curve are minimum. Thus, an array of *N* bits is expressed *1* as selected nodes and *0* as non-selected. Thus, the alternatives for each variable are [0,1]. Each location for a dolphin echolocation is called as solution. These solutions can be illustrated as Figure 1.

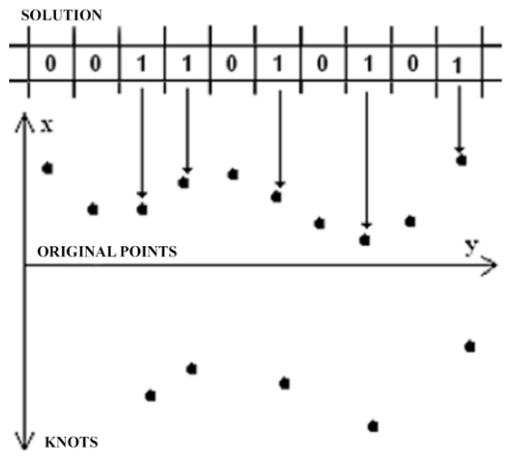

Figure 1. Sample solution illustration

For example, it is possible to express 10 points in this way with the control points to be calculated for the 5 selected nodes. The aim of dolphin echolocation is maximizing the fitness for equation (7) can be used as fitness function.

$$f = \frac{1}{Number\,of\,ControlPoint * ErrorValue} \#(7)$$

The B-Spline curve fitting process with the dolphin echolocation algorithm is as follows.

1. Create random solutions for the startup population.

2. Calculate the *PP* of current iteration.

3. Calculate the fitness value for all possible solutions.

4. Calculate the cumulative fitness of the variables in each possible solution.

5. Find the best solution according to maximum fitness.

6. Set the cumulative fitness of all solutions variables to 0 which variables equal to the variables of the best solution.

7. Calculate the probabilities of alternatives for each variable in all solutions.

8. Set the probabilities of all alternatives equal to the variables of the best solution to probability of the current iteration.

9. Find the possible solutions to be used in the next iteration by the probabilities of the alternatives for each variable.

10. Repeat steps 2-9 for the number of iteration times.

4. EXPERIMENTAL RESULTS

A. *Experimental Curve*

The target is a curve of 200 points. The approximation results of the 3rd degree B-Spline curves are as shown in Table 1.

| | GENETIC ALGORITHM | | | | DOLPHIN ECHOLOCATION ALGORITHM | | | |
|---|---|---|---|---|---|---|---|---|
| **ITERATION** | RMSE | Euclidean Distance | Number of Control Point | Fitness | RMSE | Euclidean Distance | Number of Control Point | Fitness |
| **10** | 4,36660 | 670,6340 | 92 | 61698,32 | **3,92919** | 572,6328 | 95 | **54400,11** |
| **25** | 3,90026 | 607,9586 | 96 | 58364,02 | **3,25922** | 471,9693 | 93 | **43893,14** |
| **50** | 2,18640 | 308,7266 | 108 | 33342,48 | **1,62066** | 233,4690 | 97 | **22646,50** |
| **100** | 2,93898 | 418,3969 | 93 | 38910,91 | **1,88715** | 252,2174 | 98 | **24717,30** |
| **250** | 2,49795 | 350,5132 | 97 | 33999,78 | **1,81226** | 244,5128 | 84 | **20539,08** |
| **500** | **1,75746** | 243,1649 | 101 | 24559,65 | 1,75954 | 226,2702 | 93 | **21043,13** |
| **1000** | 1,97076 | 278,8363 | 97 | 27047,12 | **1,41782** | 201,0667 | 89 | **17894,94** |
| **2000** | 1,66949 | 256,1887 | 98 | 25106,49 | **1,59712** | 201,9198 | 87 | **17567,02** |

Table 1. Experimental results for different number of iteration.
Plotted experimental results is shown in Figure 2.

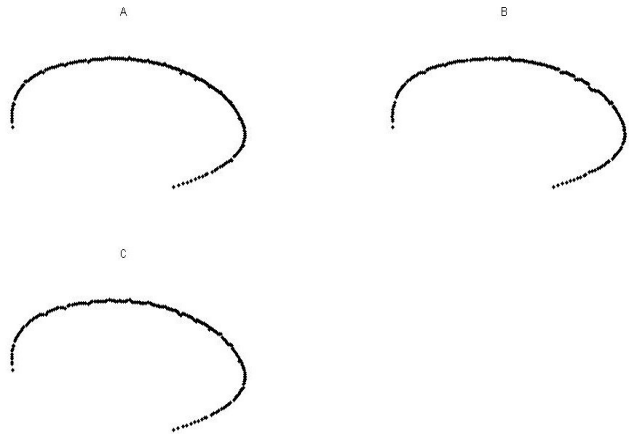

Figure 2. A) Original Curve, B) Genetic Algorithm, C) Dolphin Echolocation Algorithm

## B. Epitrochoid Curve

The target is a curve of 361 points. The curve equation is as follows.

$$x(t) = (a + b) * \cos(t) - h * \cos\left(\left(\frac{a}{b} + 1\right) * t\right)$$

$$y(t) = (a + b) * \sin(t) - h * \sin\left(\left(\frac{a}{b} + 1\right) * t\right)$$

For the parameters *a = 5*, *b = 1* and *h = 4*, The approximation results of the 3rd degree B-Spline curve for the curve calculated at *-180 <= t <= 180* are as in Table 2.

| | GENETIC ALGORITHM | | | | DOLPHIN ECHOLOCATION ALGORITHM | | | |
|---|---|---|---|---|---|---|---|---|
| **ITERATION** | RMSE | Euclidean Distance | Number of Control Point | Fitness | RMSE | Euclidean Distance | Number of Control Point | Fitness |
| **10** | 0,01510 | 5,2401 | 174 | 911,78 | **0,00689** | 3,3247 | 189 | **628,37** |
| **25** | **0,00795** | 3,1472 | 196 | **616,84** | 0,00851 | 3,4572 | 192 | 663,78 |
| **50** | 0,00937 | 3,6672 | 179 | **656,44** | **0,00906** | 3,6758 | 185 | 680,02 |
| **100** | **0,00460** | 2,4077 | 193 | **464,69** | 0,00652 | 2,9933 | 195 | 583,69 |
| **250** | 0,00767 | 3,2174 | 164 | **527,65** | **0,00739** | 2,9937 | 185 | 553,83 |
| **500** | **0,00571** | 2,5488 | 170 | **433,30** | 0,00638 | 3,1269 | 181 | 565,97 |
| **1000** | **0,00245** | 1,7965 | 189 | 339,54 | 0,00252 | 1,7468 | 179 | **312,68** |
| **2000** | 0,00314 | 1,8436 | 198 | 365,03 | **0,00308** | 1,9796 | 166 | **328,61** |

Table 2. Experimental results for different number of iteration.

Plotted experimental results is shown in Figure 3.

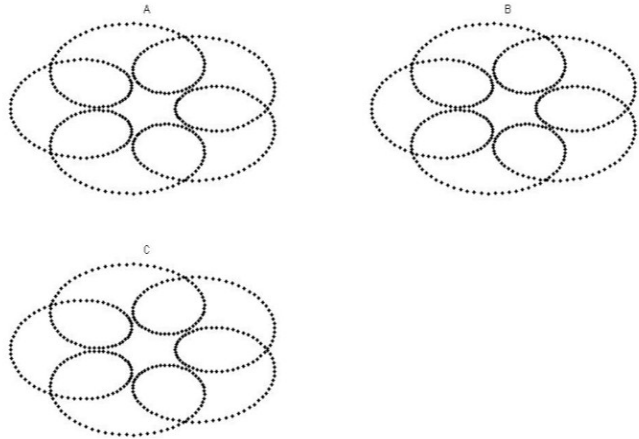

Figure 3. A) Original Curve, B) Genetic Algorithm, C) Dolphin Echolocation Algorithm

*C. Archimedean Spiral*

The target is a curve of 100 points. The curve equation is as follows.
$$r = a * t$$
$$x(t) = r * \cos(t)$$
$$y(t) = r * \sin(t)$$
For the *a = 2*, The approximation results of the 3rd degree B-Spline curve for the curve calculated at *0 <= t <= π* are as shown in Table 3.

| | GENETIC ALGORITHM | | | | DOLPHIN ECHOLOCATION ALGORITHM | | | |
|---|---|---|---|---|---|---|---|---|
| **ITERATION** | RMSE | Euclidean Distance | Number of Control Point | Fitness | RMSE | Euclidean Distance | Number of Control Point | Fitness |
| **10** | 0,00468 | 1,3419 | 46 | 61,73 | **0,00279** | 1,2520 | 47 | **58,84** |
| **25** | **0,00356** | 1,2649 | 43 | **54,39** | 0,00468 | 1,3680 | 48 | 65,67 |
| **50** | **0,00150** | 1,1490 | 50 | **57,45** | 0,00417 | 1,3182 | 44 | 58,00 |
| **100** | **0,00178** | 1,1159 | 48 | 53,56 | 0,00340 | 1,2512 | 38 | **47,54** |
| **250** | **0,00209** | 1,1604 | 38 | 44,09 | 0,00535 | 1,3127 | 32 | **42,01** |
| **500** | **0,00205** | 1,1405 | 38 | 43,34 | 0,00313 | 1,2247 | 34 | **41,64** |
| **1000** | **0,00209** | 1,1435 | 41 | 46,88 | 0,00257 | 1,2135 | 36 | **43,69** |
| **2000** | 0,00318 | 1,2054 | 33 | **39,78** | **0,00277** | 1,1890 | 36 | 42,81 |

Table 3. Experimental results for different number of iteration.

Plotted experimental results is shown in Figure 4.

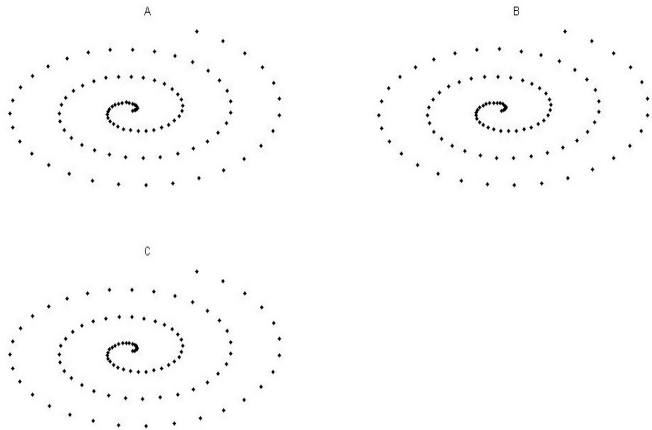

Figure 4. A) Original Curve, B) Genetic Algorithm, C) Dolphin Echolocation Algorithm

### D. Vivaldi Curve

The target curve is a curve of 241 points. The curve equation is as follows.

$$x(t) = a * (1 + \cos(t))$$
$$y(t) = a * \sin(t)$$
$$z(t) = 2 * a * \sin\left(\frac{1}{2} * t\right)$$

For *a = 0.5*, The approximation results of the 3rd degree B-Spline curve for the curve calculated at *-360 <= t <= 360* are as shown in Table 4.

| | GENETIC ALGORITHM | | | | DOLPHIN ECHOLOCATION ALGORITHM | | | |
|---|---|---|---|---|---|---|---|---|
| ITERATION | RMSE | Euclidean Distance | Number of Control Point | Fitness | RMSE | Euclidean Distance | Number of Control Point | Fitness |
| 10 | 0,00036 | 1,0711 | 108 | 115,67 | **0,00026** | 1,0599 | 92 | **97,52** |
| 25 | **0,00021** | 1,0578 | 101 | 106,84 | 0,00029 | 1,0636 | 96 | **102,11** |
| 50 | **0,00024** | 1,0431 | 90 | 93,88 | 0,00054 | 1,0957 | 77 | **84,37** |
| 100 | **0,00015** | 1,0271 | 96 | 98,60 | 0,00037 | 1,0805 | 79 | **85,36** |
| 250 | **0,00013** | 1,0231 | 95 | 97,20 | 0,00062 | 1,1046 | 65 | **71,80** |
| 500 | **0,00015** | 1,0256 | 80 | 82,05 | 0,00074 | 1,1149 | 53 | **59,09** |
| 1000 | **0,00018** | 1,0327 | 71 | 73,32 | 0,00036 | 1,0759 | 54 | **58,10** |
| 2000 | **0,00028** | 1,0419 | 59 | 61,47 | 0,00066 | 1,1431 | 40 | **45,72** |

Table 4. Experimental results for different number of iteration.

Plotted experimental results is shown in Figure 5.

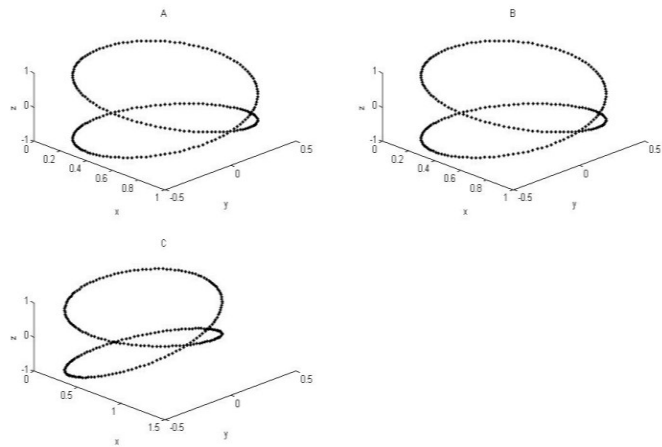

Figure 5. A) Original Curve, B) Genetic Algorithm, C) Dolphin Echolocation Algorithm

5. CONCLUSION AND FEATURE WORK

This paper addresses the problem of curve fitting of noisy data points by using B-spline curves. Given a set of noisy data points, the goal is to compute all parameters of the approximating polynomial B-spline curve that best fits the set of data points in the least-squares sense. This is a very difficult, overdetermined, continuous, multimodal, and multivariate nonlinear optimization problem. Our proposed method solves it by applying the dolphin echolocation algorithm. Our experimental results show that the presented method performs very well by fitting the data points with a high degree of accuracy. A comparison with the most popular previous approach genetic algorithm to this problem is also carried out. It shows that our method outperforms previous approaches for the examples discussed in this paper. Future work includes the extension of this method to other families of curves, such as NURBS and the parametric B-spline curves. The extension of these results to the case of explicit surfaces is also part of our future work.